\documentclass[fleqn,10pt]{wlscirep}
\usepackage[utf8]{inputenc}
\usepackage[T1]{fontenc}
\usepackage{amsfonts}
\usepackage{booktabs}
\usepackage{siunitx}
\usepackage{multirow}
\usepackage{listings}

\title{Modulating human brain responses via optimal natural image selection and synthetic image generation}

\author[1]{Zijin Gu}
\author[2]{Keith Jamison}
\author[1,2]{Mert R. Sabuncu}
\author[2,*]{Amy Kuceyeski}
\affil[1]{School of Electrical and Computer Engineering, Cornell University and Cornell Tech, New York, NY, USA}
\affil[2]{Department of Radiology, Weill Cornell Medicine, New York, New York, USA}

\affil[*]{amk2012@med.cornell.edu}

\begin{abstract}
One of the main goals of neuroscience is to understand how biological brains interpret and process incoming environmental information. Building computational encoding models that map images to neural responses is one way to pursue this goal. Moreover, generating or selecting visual stimuli designed to achieve specific patterns of responses allows exploration and control of neuronal firing  rates or regional brain activity responses. 
Here, we investigated the brain's regional activation selectivity and inter-individual differences in human brain responses to various sets of natural and synthetic (generated) images via two functional MRI (fMRI) studies. 
For our first fMRI study, we used a pre-trained group-level neural model for selecting or synthesizing images that are predicted to maximally activate targeted brain regions. We then presented these images to subjects while collecting their fMRI data. Our results show that optimized images indeed evoke larger magnitude responses than other images predicted to achieve average levels of activation.Furthermore, the activation gain is positively associated with the encoding model accuracy. While most regions' activations in response to maximal natural images and maximal synthetic images were not different, two regions, namely anterior temporal lobe faces (aTLfaces) and fusiform body area 1 (FBA1), had significantly higher activation in response to maximal synthetic images compared to maximal natural images.
On the other hand, three regions; medial temporal lobe face area (mTLfaces), ventral word form area 1 (VWFA1) and ventral word form area 2 (VWFA2), had higher activation in response to maximal natural images compared to maximal synthetic images. In our second fMRI experiment, we focused on probing inter-individual differences in face regions' responses and found that individual-specific synthetic (and not natural) images derived using a personalized encoding model elicited significantly higher responses compared to synthetic images derived from the group-level or other subjects' encoding models. Finally, we replicated the finding showing synthetic images elicited larger activation responses in the aTLfaces region compared to natural image responses in that region. Here, for the first time, we leverage our data-driven and generative modeling framework NeuroGen to probe inter-individual differences in and functional specialization of the human visual system. Our results indicate that NeuroGen can be used to modulate macro-scale brain regions in specific individuals using synthetically generated visual stimuli.

\end{abstract}
\begin{document}

\flushbottom
\maketitle

\thispagestyle{empty}

\section*{Introduction}
The brain's visual system has long been a topic of neuroscientific study, with some of the earliest classic psychological models for object recognition being performed over 100 years ago. The identification of preferences in the response patterns of single neurons \cite{hubel1962receptive, hubel1968receptive} and macro-scale regions  \cite{kanwisher1997fusiform, epstein1998cortical, downing2001cortical} in the visual cortex has enabled understanding of how brains process and interpret incoming visual information. Artificial neural networks (ANNs), especially deep neural networks, with their architecture motivated by biological neural networks (BNNs) and their striking performance on image classification and object recognition tasks, have naturally lead to their use in modeling the human visual system. Recent work has specifically focused on comparing ANNs trained to predict brain responses from visual stimuli, called encoding models, to the brain's visual system \cite{kubilius2016deep, kubilius2019brain, zhuang2021unsupervised, mehrer2021ecologically, sexton2022reassessing, Schrimpf2020integrative}. For instance, Kubilius et al. developed a shallow, recurrent ANN representing anatomical brain structures that was shown to accurately reproduce the flow of activity in the primate ventral visual stream \cite{kubilius2019brain}. Zhuang et al. found unsupervised ANNs can produce brain-like representations and can achieve accuracy in predicting cortical activations that equals or exceeds best supervised methods \cite{zhuang2021unsupervised}; similarly, Mehrer et al. showed ANNs trained on a dataset of brain responses from 1.5 million images across 565 basic categories  better predicted representations in higher-level human visual cortex and perceptual judgments than on typical image classification dataset, e.g. ImageNet \cite{mehrer2021ecologically}. While some recent work has highlighted mismatches between ANNs and BNNs \cite{bowers2022deep}, ANNs remain among the best models for representing and probing visual systems.   

Due in part to recent AI breakthroughs in generative models, e.g., generative adversarial networks (GANs) \cite{brock2018large}, variational autoencoders (VAEs) \cite{van2017neural} and diffusion models \cite{rombach2022high}, neural decoding and optimal stimulus design have gained popularity as novel ways to understand and control neural responses to visual stimuli. Coupling pretrained generators with linear or ANN-based encoding models has allowed accurate decoding of viewed images from brain responses that have both high-level semantic and low-level alignment with ground truth \cite{ozcelik2022reconstruction,gu2022decoding}. ANNs that perform image classification can be coupled with generative networks to synthesize preferred inputs for artificial neurons via activation maximization \cite{nguyen2016synthesizing}. Neuroscience researchers have adopted similar approaches for designing optimal stimuli for maximizing firing rate in single neurons or populations of neurons in macaque monkeys \cite{bashivan2019neural, ponce2019evolving}. Bashivan et al. showed that the firing rate of V4 neural sites can be controlled by a deep ANN as a group, and, to some extent, independently \cite{bashivan2019neural}. Ponce et al. revealed the response properties of visual neurons in V1 by exploring the vast generative image space \cite{ponce2019evolving}. In terms of human studies, previous work with GAN-based image synthesis showed promising results in testing category selectivity of brain regions and discovering inter- individual and regional difference \cite{ratan2021computational,gu2022neurogen}. However, to the best of our knowledge, there is no work thus far that has recorded macro-scale human brain activation in response to synthetic visual stimuli designed to achieve specific, targeted brain activation patterns.

In this work, we aim to enrich our understanding of the human visual system by attempting to modulate activation responses in specific regions of the human brain using selected natural and specifically designed synthetic visual stimuli. We used the large-scale Natural Scences Dataset (NSD) \cite{allen2021massive}, consisting of ~30-40K coupled images and brain responses from each of 8 subjects, to train individual-level ANN-based encoding models with high accuracy. By feeding the NSD images into these encoding models and sorting their predicted average activations, we obtained sets of natural images that were predicted to achieve maximal (or average) levels of activity for that region across the population of NSD subjects. In addition, we used the previously developed NeuroGen framework to design synthetic images predicted to achieve the same goals \cite{gu2022neurogen}. Once the natural and synthetic image sets were obtained, we prospectively enrolled six novel individuals and measured their brain responses to these images via fMRI. Once we had image-response data from the six prospectively enrolled subjects, we applied our recently developed linear ensemble method to create personalized, individual-level encoding models for each of these new subjects \cite{gu2022personalized}. We then used their personalized encoding models for a particular face region to obtain sets of individual-specific natural and synthetic images via the same image selection/generation procedure described in the first experiment, and obtained their regional responses to these personalized images during a second fMRI scan. We demonstrate that the proposed method can be used to select and generate optimal visual stimuli designed to modulate macro-scale human brain activity in a targeted manner, and, further, that this modulation can be done at the level of a specific individual.

\section*{Results}
\begin{figure}[htbp]
    \centering
    \includegraphics[width=\linewidth]{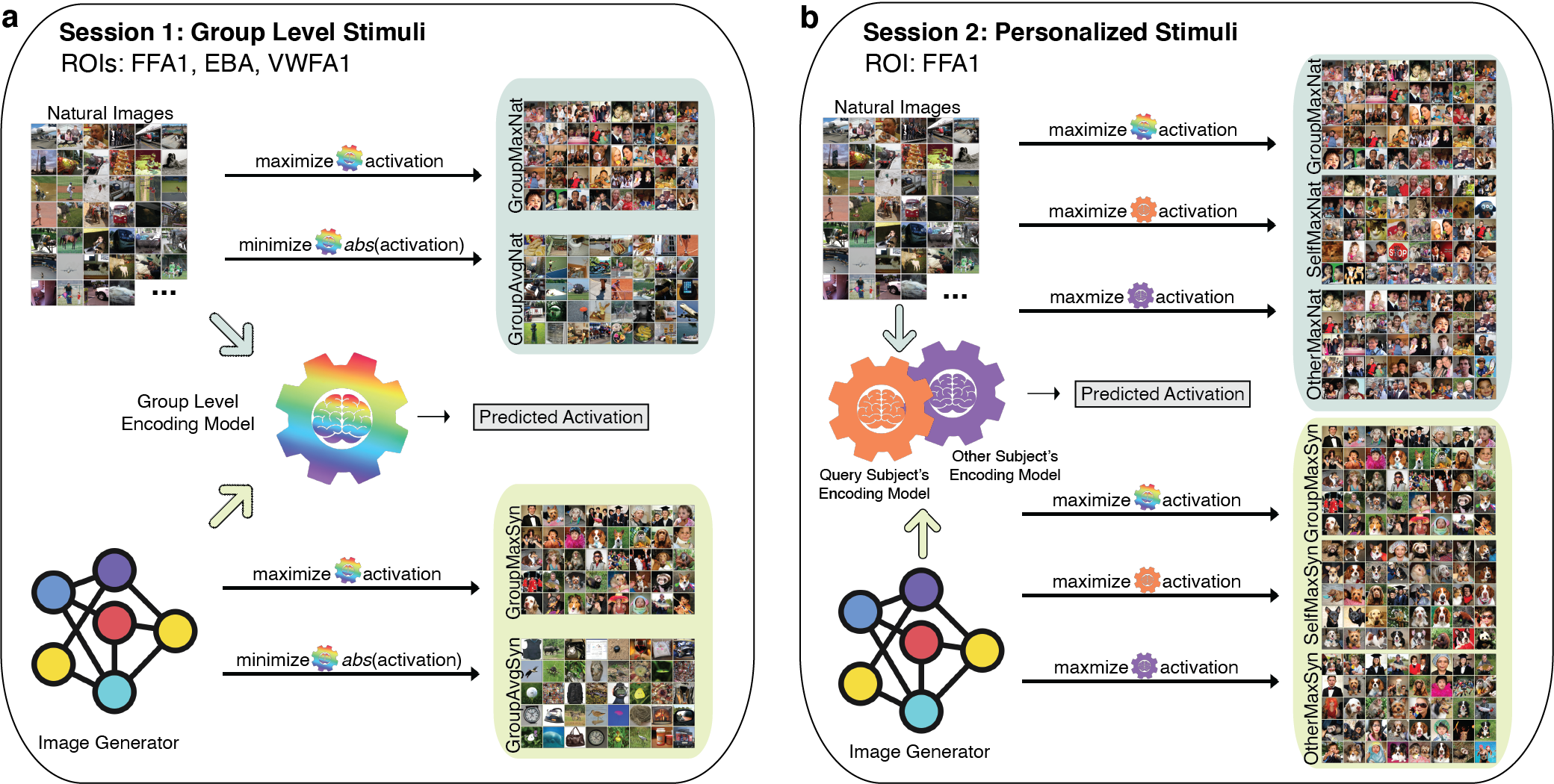}
    \caption{Experimental design workflow. \textbf{a} Session 1: The first experiment involved showing four sets of natural and synthetic images to each of 6 subjects while they underwent fMRI. These sets of images were selected based on their predicted average activation across the 8 NSD subject-level encoding models ("Group" model). Candidate natural images were the set of 9,000 $\times$ 8 = 72,000 images shown to any subject in the NSD experiments, excepting the shared 1000 images ("Nat"). Synthetic images were created using NeuroGen, which uses BigGAN-deep as its generator ("Syn"). The first set of images, called "GroupMaxNat", are the natural images with the highest  predicted activation in the group NSD encoding model. The second set of images, called "GroupAvgNat", are the natural images with predicted activations in the NSD group model that are closest to average. The final two sets of images, called "GroupMaxSyn" and "GroupAvgSyn", are synthetic images designed by NeuroGen to achieve maximal and average activation in the group NSD encoding model, respectively. The regions of interest, or targets, for the session 1 experiments are FFA1, EBA and VWFA1. \textbf{b} Session 2: Session 1 data was used to create a personalized encoding model for each of the six subjects, and these personalized encoding models ("Self") were used to select natural and generate synthetic images that were predicted to achieve maximal activation for that person's FFA1 encoding model, named as "SelfMaxNat" and "SelfMaxSyn". During Session 2, we also showed each subject Session 1's group maximal images ("GroupMaxNat" and "GroupMaxSyn") and the other subjects' personalized images ("OtherMaxNat" and "OtherMaxSyn") for FFA1 to test the specificity of the personalization. Face regions: OFA - occipital face area; FFA - fusiform face area; mTLfaces - medial temporal lobe face area; aTLfaces - anterior temporal lobe face area. Body regions: EBA - extrastriate body area; FBA - fusiform body area;  mTLbodies - medial temporal lobe body area. Word regions: OWFA - occipital word form area; VWFA - visual word form area; mfswords - mid-fusiform sulcus word area; mTLwords - medial temporal lobe word area.} 
    \label{fig:workflow}
\end{figure}
Figure \ref{fig:workflow} shows our workflow consisting of the natural image selection and synthetic image generation process for Session 1 and Session 2's fMRI experiments. Three visual regions, each from a different perception group, i.e. fusiform face area 1 (FFA1), extrastriate body area (EBA) and visual word form area 1 (VWFA1), were identified as primary targets. We began by creating 8 individual-level encoding models for these three regions using the image-response data from NSD and the deepnet feature-weighted receptive field (deepnet-fwRF) model described in previous work \cite{st2018feature}. A simple average of the 8 individual-level encoding models were taken to obtain a "Group" average encoding model. Our set of candidate natural images were the 9,000 $\times$ 8 = 72,000 images shown to any one of the 8 NSD subjects that did not belong to the shared 1000 image set. From this set, we selected the top 40 natural images ("Nat") that maximized a region's predicted activation ("Max") for the "Group" encoding model (called “GroupMaxNat”) and the 40 natural images ("Nat") that minimized a region's absolute predicted activation ("Avg") for the "Group" encoding model (called “GroupAvgNat”). We also inserted the "Group"  encoding model into the NeuroGen framework \cite{gu2022neurogen} and generated 40 synthetic images ("Syn") that maximized a region's predicted activation (called “GroupMaxSyn”) and 40 synthetic images that minimized a region's absolute predicted activation (called “GroupAvgSyn”). Session 1's experiments consisted of six novel subjects (called NeuroGen subjects) that underwent fMRI while viewing the images in these 12 stimulus sets (4 sets $\times$ 3 regions).

Session 2's image set generation followed a similar procedure to Session 1, with the main differences being that the "Group" encoding model was replaced with individual-level, personalized encoding models (called "Self"), and that we focused only on the face region FFA1. The personalized encoding models were fit using each subject's image-FFA1 response paired data from Session 1 and our recently published linear ensemble approach \cite{gu2022personalized}. For each of the 6 NeuroGen subjects, "SelfMaxNat" or "SelfMaxSyn" are sets of natural or synthetic images that maximized FFA1's predicted activation for that subject. Besides showing the "Self" condition images in Session 2, we also showed subjects the personalized FFA1 images of other subjects (called "OtherMaxNat" and "OtherMaxSyn") and the "Max" group level FFA1 images from Session 1 ("GroupMaxNat" and "GroupMaxSyn").  Session 2's fMRI experiments consisted of recording brain responses to 128 images from a total of 6 conditions (32 "SelfMaxSyn", 32 "SelfMaxNat", 20 "OtherMaxSyn", 20 "OtherMaxNat", 12 "GroupMaxSyn" and 12 "GroupMaxNat").

\subsection*{Observed and targeted brain activation patterns are well aligned}
\begin{figure}[htbp]
    \centering
    \includegraphics[width=\linewidth]{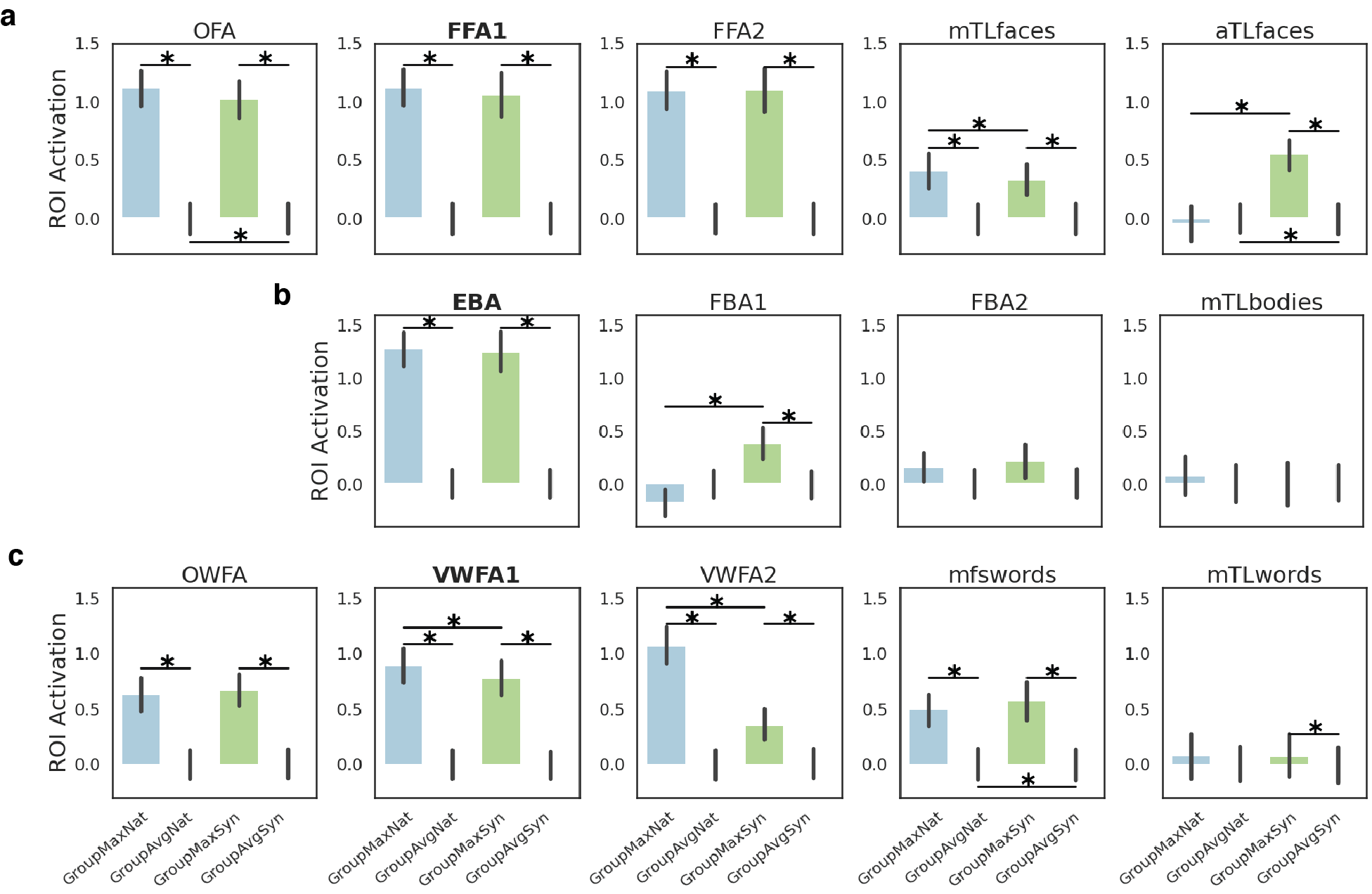}
    \caption{Brain activations in response to the "Max" condition images are significantly higher than the activations in response to the "Avg" condition images for their targeted region (and for almost all regions in the same response category), for both natural ("Nat") and synthetic ("Syn") image sets. Barplots show the mean and standard deviation of normalized fMRI activity in response to the different image conditions in \textbf{a} face regions, \textbf{b} body regions and \textbf{c} word regions. The activity normalization was done with respect to the activity from "Avg" condition images of the same source. The following group comparisons were performed: "GroupMaxNat" vs "GroupAvgNat", "GroupMaxSyn" vs "GroupAvgSyn", "GroupMaxSyn" vs "GroupMaxNat" and "GroupAvgSyn" vs "GroupAvgNat". Significant differences based on permutation testing ($p < 0.05$) are marked with a starred horizontal line. }
    \label{fig:session1-group-comp}
\end{figure}

We fit a linear mixed effects (LME) model and compared the brain activations between image conditions ("GroupMaxNat", "GroupAvgNat", "GroupMaxSyn", "GroupAvgSyn") for all six subjects in the three primary target regions, see Figure \ref{fig:session1-group-comp}a. We found that "Max" images had significantly higher activity compared to "Avg" images from the same source (natural or synthetic) and no significant differences were found between the "GroupMaxSyn" and "GroupMaxNat" activation responses for FFA1 and EBA, but the natural images have significantly higher responses in VWFA1. Though inter-regional differences exist, regions belonging to the same perception group (see Figure \ref{fig:session2-lme}a and Figure S2 for the anatomical location of the regions) are usually activated by similar features. Thus, we analyzed all visual regions in the same perception category for stimuli designed for the primary target region from that category. Figure \ref{fig:session1-group-comp}a shows other face regions' (OFA, FFA1, FFA2, mTLfaces and aTLfaces) activations in response to images designed for FFA1, body regions' (EBA, FBA1, FBA2 and mTLbodies) activations in response to images designed for EBA, and word regions' (OWFA, VWFA1, VWFA2, mfswords and mTLwords) activations in response to images designed for VWFA1. Generally, we observed significantly larger activation in secondary target regions in response to maximal images compared to average images, except in aTLfaces, FBA1, and mTLwords for the natural images and FBA2 and mTLbodies for both natural and synthetic images. Finally, maximal synthetic images achieved significantly higher activations than maximal natural images in secondary target regions aTLfaces and FBA1, while maximal natural images achieved significantly higher activations than maximal synthetic images in the primary word target region VWFA1 and secondary target regions mTLfaces and VWFA2. 

\subsection*{More accurate encoding models result in better alignment of observed and targeted brain activation patterns}
\begin{figure}[htbp] 
    \centering
    \includegraphics[width=\linewidth]{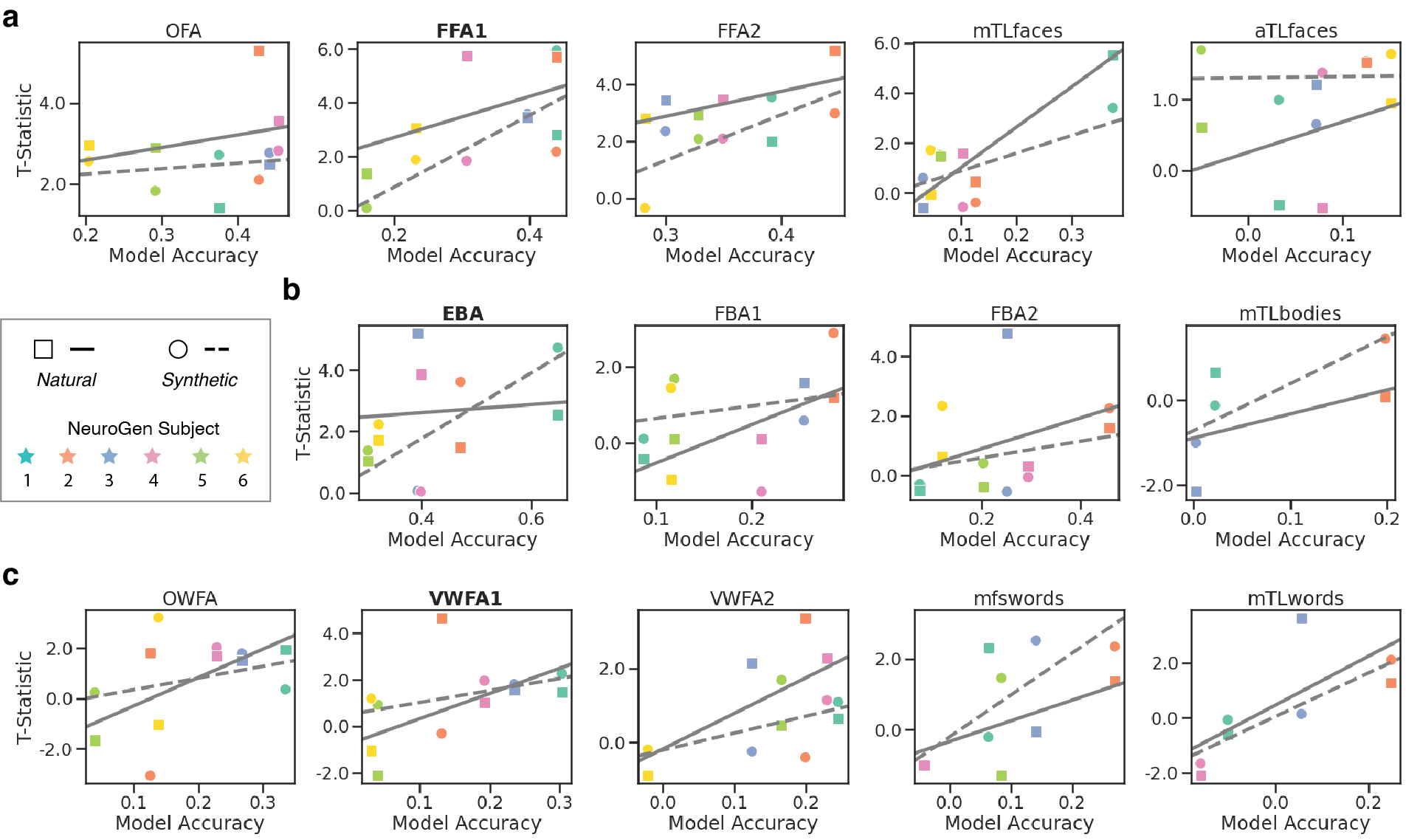}
    \caption{Positive correlations between "Max" vs "Avg" activation differences (captured via unpaired t-statistics) and encoding model accuracy across the six NeuroGen subjects, where each subject is the same color across all scatter plots. Squares indicate the "Nat" image results where the line of best fit is drawn with a solid line and circles indicate the "Syn" image results where the line of best fit is drawn with a dashed line. Note: some regions were not identifiable in some subjects using the localizer scans. Scatter plots \textbf{a}, \textbf{b} and \textbf{c} show the relationship between encoding model accuracy (x-axis) and the t-statistic calculated by contrasting "GroupMaxNat" vs "GroupAvgNat" or "GroupMaxSyn" vs "GroupAvgSyn" for face, body and word perception regions respectively.}
    \label{fig:session1-acc-t}
\end{figure}

We hypothesize that the success of our natural or synthetic "Max" images in driving brain activity higher than activity in response to natural or synthetic "Avg" images hinges on the accuracy of the "Group" encoding model for that subject. First, we found that while the "Group" encoding model had a trend toward better accuracy for the natural image responses compared to the synthetic image responses (t-statistic = 1.5, p = 0.11), overall the prediction accuracies were similar for the two image types (Pearson's $r = 0.467, p = 1.652e-5$), see Figure S4. To test our hypothesis about the relationship between accuracy and success in modulation, we correlated subjects’ encoding model accuracy value and the t-statistic representing activation differences in brain responses to "Max" and "Avg" conditions for natural and synthetic images, see Figure \ref{fig:session1-acc-t}a ,b, and c for face, body and word perception groups respectively. We found that these correlations were indeed all generally moderate to high positive values, with overall $p$-value < 0.0001 for synthetic stimuli and 0 for natural stimuli, based on permutation test. 

\begin{figure}[htbp] 
    \centering
    \includegraphics[width=0.9\linewidth]{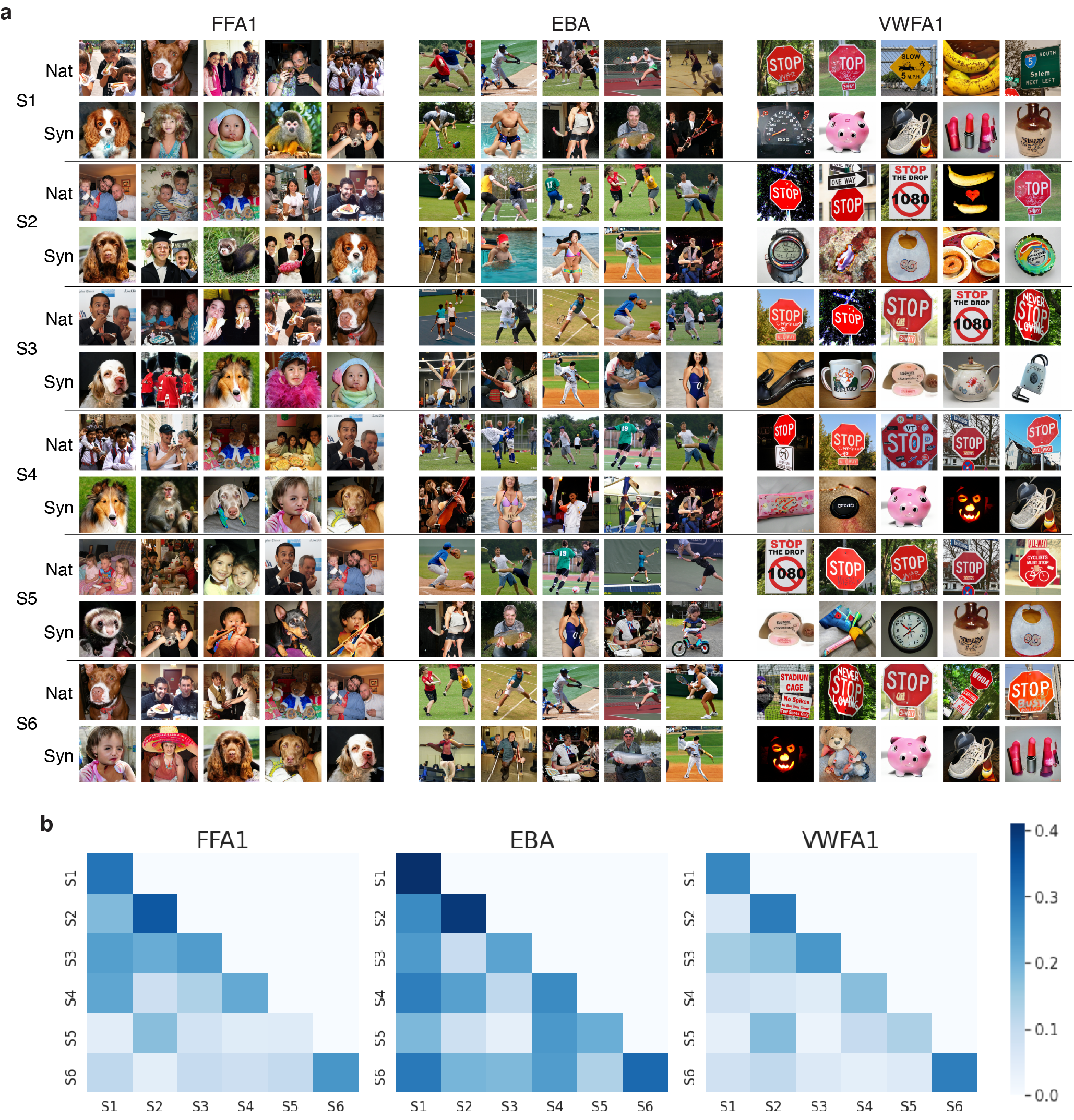}
    \caption{\textbf{a} For each of the six NeuroGen subjects, the 5 synthetic and 5 natural images that had the highest observed activation in FFA1, EBA and VWFA1. \textbf{b} Off-diagonal elements quantify individual differences via across-subject correlation of FFA1, EBA and VWFA1 responses to the same stimulus image, while the diagonal elements show within-subject reliability calculated by correlating FFA1, EBA and VWFA1 responses to two presentations of the same stimulus image. }
    \label{fig:session1-ind-diff}
\end{figure}

In Figure \ref{fig:session1-ind-diff}a, we display each of the six subject's 5 most activating natural and synthetic images for FFA1, EBA and VWFA1, sorted in descending order by their measured Session 1 fMRI responses. We observed that while a few of the same top images appear across different subjects, most are not shared across individuals. In Figure \ref{fig:session1-ind-diff}b, we performed a pair-wise correlation of each subjects' brain activity responses to quantify inter-subject similarity and found that subjects' responses to the same image vary quite widely, with across-subject correlations ranging from 0 to 0.35. To compare against the noise ceiling, we also include in the diagonal of Figure \ref{fig:session1-ind-diff}b the test-retest within-subject reliability in responses to the same image. We see that, in most cases, the diagonal value is larger than the off-diagonal entries.

\subsection*{Personalized synthetic images allow probing individual differences in brain responses}
\begin{figure}[htbp]
    \centering
    \includegraphics[width=0.8\linewidth]{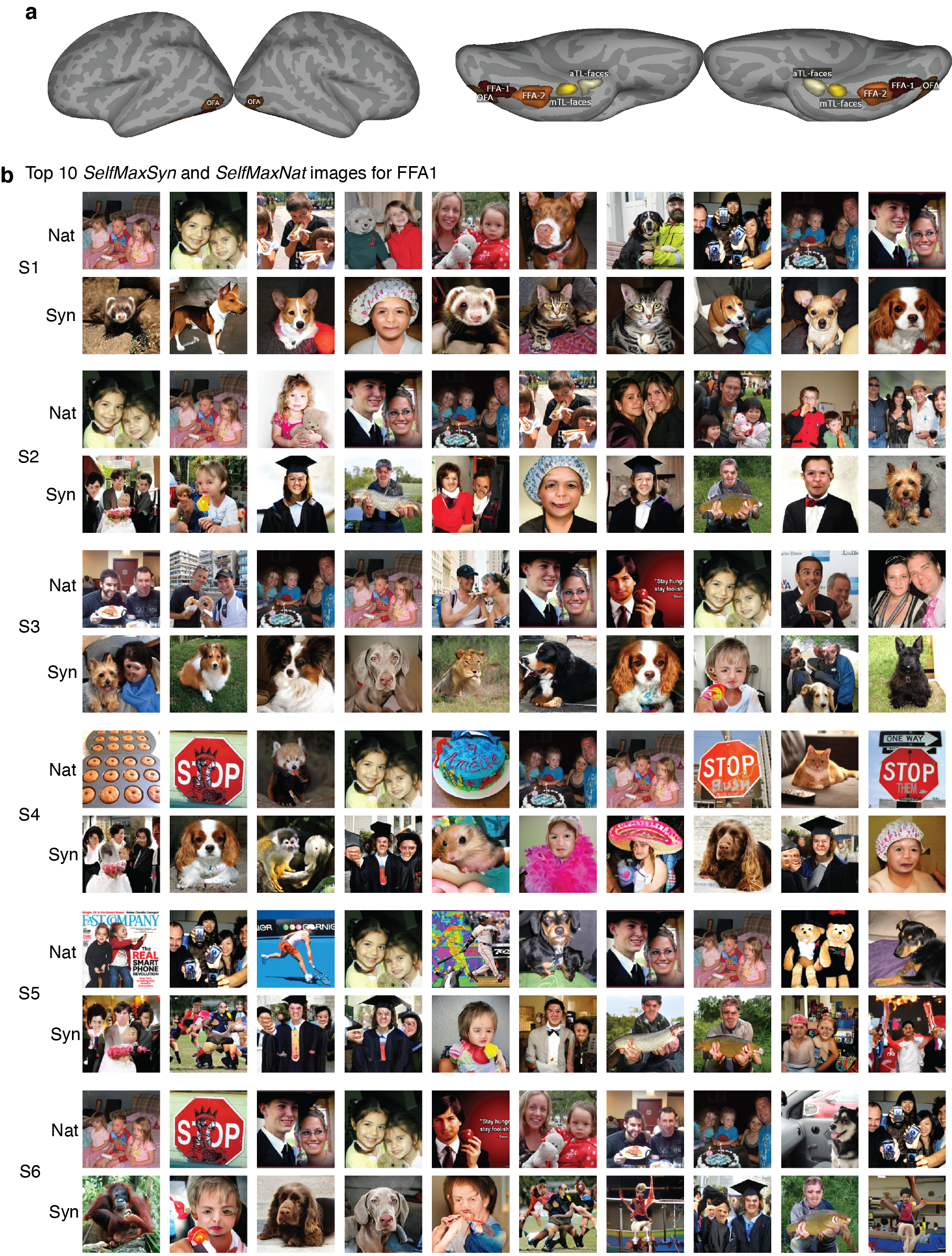}
    \caption{The effect of personalization in achieving targeted brain activation patterns in face perception areas. \textbf{a} The anatomical location of the five face regions in the visual cortex. \textbf{b} For each of the six subjects, the ten synthetic and ten natural personalized images that elicited the highest observed FFA1 responses according to the fMRI measurements, in descending order. }
    \label{fig:session2-lme}
\end{figure}
We have shown that selecting and generating images using a group-level encoding model allows targeted modulation of regional brain activity in prospective, novel individuals. Given that there are individual differences in brain responses to images, we hypothesized that selecting and generating personalized natural and synthetic images using an individual-level encoding model might allow more enhanced modulation of regional brain responses. We focused on FFA1, as the face perception regions showed consistent and promising results in group level analysis, see Figure \ref{fig:session1-group-comp}a. We trained individual-level linear ensemble encoding models for each of our 6 novel subjects' FFA1 regions using data from their Session 1 experiments (see details in the Method section). Personalized natural and synthetic images were selected or generated based on the personalized encoding models ("Self") in the same way as described previously to obtain "SelfMaxNat" and "SelfMaxSyn" images for FFA1. To test if there was a boost in activation responses from the personalization compared to the group-level images, we fit LME models to compare "GroupMaxSyn" vs "SelfMaxSyn" and "GroupMaxNat" vs "SelfMaxNat". To test the inter-individual specificity of the personalization, we fit LME models to compare "OtherMaxSyn" vs "SelfMaxSyn" and "OtherMaxNat" vs "SelfMaxNat", where "Other" indicates those images are a random subset of the other subjects' personalized image sets. Finally, we tested if the personalized synthetic images had response activations that were higher than the personalized natural images with LME models comparing "SelfMaxNat" vs "SelfMaxSyn". We performed these comparisons for all regions in the face perception category; Figure \ref{fig:session2-lme}a shows the anatomical locations of the 5 face perception regions of interest.

\begin{table}[htbp]
\centering 
\begin{tabular}{c|cc|cc|cc|cc|cc}
\toprule
\multirow{2}{*}{contrast} & \multicolumn{2}{c|}{OFA} & \multicolumn{2}{c|}{\textbf{FFA1}} & \multicolumn{2}{c|}{FFA2} & \multicolumn{2}{c|}{mTLfaces} & \multicolumn{2}{c}{aTLfaces}\\ \cmidrule{2-11} 
                          & $\beta$ & $p$-value & $\beta$ & $p$-value      & $\beta$ & $p$-value     & $\beta$ & $p$-value & $\beta$ & $p$-value         \\ \midrule
GroupMaxSyn vs SelfMaxSyn     & -0.016 &0.718&-0.002&0.984& 0.076 &0.057& 0.214 &0.010*& 0.122 &0.046*\\ \midrule
OtherMaxSyn vs SelfMaxSyn  & 0.019 &0.614& 0.074 &0.099&  0.112 & 0.001*& 0.179 &0.004*& 0.134 & 0.007*\\ \midrule
GroupMaxNat vs SelfMaxNat & -0.034 &0.419&  0.039 & 0.449& 0.008 &0.836&  0.042 & 0.429& -0.136 &0.015*\\ \midrule
OtherMaxNat vs SelfMaxNat  & -0.002 &0.974& 0.048&0.262& 0.054 &0.144& 0.076 &0.122& -0.06 & 0.195\\ \midrule
SelfMaxNat vs SelfMaxSyn  &  -0.02 &0.550& -0.045 &0.239 & -0.077 &0.013*& -0.005 &0.909 &  0.103 &0.014*\\ \bottomrule
\end{tabular}
\caption{The results of fit linear mixed effects (LME) models for assessing personalization effects in face perception regions by comparing: "GroupMaxSyn" vs "SelfMaxSyn", "OtherMaxSyn" vs "SelfMaxSyn", "GroupMaxNat" vs "SelfMaxNat", "OtherMaxNat" vs "SelfMaxNat",  and "SelfMaxNat" vs "SelfMaxSyn". Columns representing different regions are arranged from posterior (lower-order) to anterior (higher-order). The $\beta$ values are the model coefficients, where positive values mean the fMRI activation from the second condition is higher than activation from the first condition; corresponding two-tailed $p$-values were calculated based on permutation testing. *uncorrected, permutation-based $p<0.05$.}
\label{tab:lme_face}
\end{table}

Figure \ref{fig:session2-lme}b shows, for each of the six subjects, the ten natural and ten synthetic images that had the highest observed responses in descending order. The image group comparison results are shown in Table \ref{tab:lme_face} via $\beta$ coefficients from the LME model, where positive means the observed responses to images in the second condition were higher than the observed responses to images in the first condition (and for negative $\beta$, vice versa). Overall, the personalization seemed to work better for the synthetic images compared to the natural images and for the hierarchically later face regions compared to the earlier ones. Specifically, there were trends for significantly higher FFA1 response to the personalized synthetic images compared to the other subjects' personalized synthetic images and higher FFA2 response to the personalized synthetic images compared to the group synthetic images. All three of the later face regions (FFA2, mTLfaces and aTLfaces) had significantly higher (permutation-based $p \lesssim 0.05$) activation in response to the personalized synthetic images compared to the other subjects' personalized images, while mTLfaces and aTLfaces also had significantly higher responses to the personalized synthetic images compared to the group synthetic images. The personalization largely had no effect in the natural image responses, with the only significant difference being that the group-level images actually had higher responses than the personalized images for mTLfaces. Finally, while most face regions' responses are not different between the personalized natural and synthetic image sets, similar to what was was found using "Group" images, we observe significantly larger responses to synthetic images compared to natural images in the highest-order face region, aTLfaces, and significantly higher responses to natural images compared to synthetic images in the mid-level FFA2 region. 

\section*{Discussion}
Inspired by previous work in macaques where neuronal firing rates could be driven by optimally designing synthetic images \cite{bashivan2019neural, ponce2019evolving}, here we carried out a set of experiments to show that human brain responses can be modulated in a controlled, personalized way by both selecting optimal natural stimuli and generating optimal synthetic stimuli. All sets of natural or synthetic images designed to maximize activity in targeted brain regions ("Max" conditions) were able to elicit significantly higher observed activity compared to images designed to achieve average activity ("Avg" conditions). Two visual regions, FBA1 and aTLfaces, had significantly higher activation in response to the maximal synthetic images compared to the activation in response to the maximal natural images, while a face area mTLfaces and two word regions VWFA1 and VWFA2 had higher activity in response to natural compared to synthetic regions. We also found that the modulation ability, quantified by the magnitude of the t-statistic between maximal and average brain responses, was associated with the accuracy of the encoding models. That is, more accurate encoding models led to more precise control over brain activity. In addition, inter-individual variability of responses in face regions was considered when creating/selecting the optimal images. We showed that personalization did indeed drive responses for specific individuals above and beyond the responses to images designed using a group-level encoding model or other individuals' encoding models, but only for synthetic images and only in face regions that were higher in the processing hierarchy. Finally, we observed that, as in Session 1's results, optimal personalized synthetic images had larger responses in the highest-level face processing region aTLfaces, compared to that regions' responses to optimal natural images. 

Classically, identifying functional specialization in the brain requires a subject to view a set of images selected by the experimenter based on a priori information or a specific hypothesis of the region of interest's preferred image characteristics \cite{saxe2006divide, friston2006critique}. This type of approach has resulted in identification of various specialized regions in the visual system, including regions that preferentially activate in response to faces \cite{kanwisher1997fusiform}, bodies \cite{downing2001cortical}, places \cite{epstein1998cortical}, objects \cite{grill2001lateral} and words \cite{cohen2002language}. However, there are clear limitations to this type of approach in that the content and characteristics of the images are selected by an experimenter with narrow focus, a highly circumscribed hypothesis and limited resources for experimentation. With the monumental progress in generative AI, the publication of large data sets containing image-response information, and improvements in encoding model accuracy using deep learning, the field can and should shift toward using a data-driven approach to selecting and designing optimal stimuli for discovery of functional specialization in the human visual system. This work takes a first step in that direction by robustly demonstrating the ability to drive responses in various human brain regions in a totally unsupervised way using "optimal" selected natural and synthetically generated images, and takes a second step in that direction by showing that personalization of image-response encoding models can allow generation of individual-specific "optimal" images (but perhaps only from synthetic sources).

While ANN-based encoding models may be improved by making them more brain-like \cite{kubilius2019brain}, the quality and size of the available individual training data is also central to encoding model accuracy. Typically, encoding models require tens of thousands of image-response pairs to obtain good alignment between predictions and observations, such as our NSD-based models that use over 20,000 training samples per subject. Following our recent work \cite{gu2022personalized}, here we constructed personalized encoding models using training data that was only $\sim$ 2\% of the NSD data sample size and obtained relatively good accuracy. However it is likely that more training data for a given subject (particularly synthetic image-response pairs) would increase encoding model accuracy and/or result in models that better reflect inter-individual differences. Since we found that the success of the "optimal" images in hitting their target responses was closely related to encoding model accuracy, we conjecture that increasing the individual image-response pair training sample size (and more repeats per image to increase SNR) may lead to better encoding model performance and more precise control over elicited brain responses. 

Previous work has mapped the face processing stream in visual cortex, namely with activity flowing from OFA to FFA (sometimes split into FFA1 and FFA2) to medial temporal and then anterior temporal lobe \cite{collins2014beyond}. We observed increasing "success" of the personalized synthetic images when arranging the results from low to high-order (OFA, FFA1, FFA2, mTLfaces to atLfaces), and little effect of personalization in the natural image responses. This could perhaps be explained by two phenomena: 1) more homogeneity in brain responses across the population to low-level characteristics of face images, like facial topology \cite{henriksson2015faciotopy}, and less homogeneity in brain responses to higher-order characteristics, like facial recognition \cite{barense2010medial, yang2016anterior} and 2) less inter-subject homogeneity in higher-order face regions' responses to the synthetic images (compared to the natural ones), as they are more novel (and perhaps demand more attention) for higher-order tasks like facial recognition. Interestingly, aTLfaces is the only face region that showed significantly higher responses for the synthetic images compared to the natural ones in both experiments, which might be reflecting this phenomena. Finally, there could be an effect generally of the iterative nature of the synthetic image generation that could result in more power to optimize image features that favor the specific subjects than selection from a set of predetermined, fixed natural images.

Unlike experiments in macaque monkeys where microelectrode arrays can be invasively implanted directly on the brain to record neuronal responses \cite{bashivan2019neural,ponce2019evolving}, human experiments obviously must rely on non-invasive recording techniques. While fMRI is one of the best non-invasive methods with which to measure human brain responses as this modality has high spatial resolution (compared to EEG, not microelectrode arrays) and whole-brain coverage, its main limitations are that it measures non-neuronal BOLD signals, has relatively low temporal resolution and can be subject to imaging artifacts related to acquisition and subject motion. Microelectrode arrays also have the advantage in that they allow recording of neuronal responses in real time so that their firing rates can be monitored in a closed-loop fashion during synthetic image optimization \cite{ponce2019evolving}. Real-time fMRI monitoring of brain responses to on-the-fly generated or selected images in an iterative, online optimization would be difficult. With our current configuration, we can only employ post hoc synthesis and offline optimization, which may partly explain the relatively weaker modulation ability in our human study compared with single-cell neuronal modulation studies in animals.

BigGAN-deep, state-of-the art at the time of the experimental design, is an image generator that allows a user-defined balance between fidelity and variety of the synthetic images. Here, we chose more emphasis on fidelity in return for lower variety since we wanted to begin our experiments with images as close as we could to the distribution of natural images used to train the encoding model. Even still, some of the generated images do not look entirely natural. The fact that synthetic maximal image responses in word areas were significantly lower than the responses to the maximal natural images could be attributed to the fact that BigGAN-deep was not trained to produce images with only text \cite{brock2018large}. Most of the VWFA1 synthetic images are of items that normally contain text (speedometer, bottles, watches, etc) but the actual text is not readable; the natural images on the other hand contain items with obvious, readable text, e.g. road signs. While the generated faces and bodies are also not completely life-like they are still recognizable as human faces or bodies. Future work, particularly including generative networks that create more accurate rendering of human faces and limbs, e.g. MidJourney v5 \cite{midjourney}, could further improve the performance of the synthetic images in achieving targeted brain activity. 

Taken together, we demonstrate here the possibility of modulating regional human brain responses in a controlled way using group- and individual-level encoding models coupled with either large databases of natural images or generative models that create synthetic images. It appears that achieving targeted control of human brain responses to visual stimuli hinge on three main issues: the accuracy and personalization of encoding models, the content/range of candidate natural image sets and the quality of the image generators. Future directions will focus on improvements in all of these domains, including incorporating semantic content or neuroscientific knowledge into encoding models and using more realistic generators, i.e. stable diffusion \cite{rombach2022high}. In summary, this approach provides a data-driven method to investigate functional specialization of and possibly a way to modulate regional brain activity in specific humans' brains by either selecting natural or designing synthetic optimal stimuli. We believe this work demonstrates the promise of generating optimal synthetic images, perhaps in the future using better generators, that may succeed in targeted, controlled modulation of brain activations and, in so doing, result in a better understanding of functional specialization within the human visual system.

\section*{Methods}
\subsection*{Data description}
\subsubsection*{Natural Scenes Dataset}
The individual encoding models were trained and tested on data from the Natural Scenes Dataset (NSD) \cite{allen2021massive}, which contains densely-sampled functional MRI (fMRI) data from eight participants (6 female, age 19-32 years). Each subject viewed 9,000–10,000 distinct color natural scenes with 2-3 repeats per scene over the course of 30-40 7T MRI sessions (whole-brain gradient-echo EPI, 1.8-mm iso-voxel and 1.6s TR). The images that subjects viewed (3s on and 1s off) were from the Microsoft Common Objects in Context (COCO) database \cite{lin2014microsoft} with a square crop resized to 8.4° $\times$ 8.4°. Among all images, a set of 1,000 were shared across all subjects while the remaining images for each individual were mutually exclusive across subjects. Subjects were asked to fixate centrally and perform a long-term continuous image recognition task ($\inf$-back) to encourage maintenance of attention. 

NSD data processing has been previously described \cite{allen2021massive}. Briefly, the fMRI data were pre-processed to correct for slice time differences and head motion using temporal interpolation and spatial interpolation. Then the single-trial beta weights representing the voxel-wise response to the image presented was estimated using a general linear model (GLM). There are three steps for the GLM: the first is to estimate the voxel-specific hemodynamic response functions (HRFs); the second is to apply the GLMdenoise technique \cite{charest2018glmdenoise, kay2013glmdenoise} to the single-trial GLM framework \cite{prince2022improving}; and the third is to use an efficient ridge regression \cite{rokem2020fractional} to regularize and improve the accuracy of the beta weights, which represent activation in response to the image. FreeSurfer was used to reconstruct the cortical surface, and both volume- and surface-based versions of the voxel-wise response maps were created. Data from the functional category localizer experiment (fLoc) \cite{stigliani2015temporal} was used to create contrast maps (voxel-wise t-statistics) of responses to specific object categories, and region boundaries were then manually drawn on inflated surface maps by identifying contiguous regions of high contrast in the expected cortical location, and thresholding to include all vertices with contrast $>$ 0 within that boundary. Early visual ROIs were defined manually using retinotopic mapping data on the cortical surface. Surface-defined ROIs were projected back to fill in voxels within the gray matter ribbon. Region-wise image responses were then calculated by averaging the voxel-wise beta response maps over all voxels within a given region.

\subsubsection*{NeuroGen Dataset}
We collected prospective data from six individuals (5 female, age 19-25) over two scans on a 3T GE-MR750 scanner, see Figure \ref{fig:workflow}. The first MRI scan included an anatomical T1 (0.9 mm iso-voxel), a functional category localizer (floc) to identify higher-order visual region boundaries (as in the NSD experiments), and, finally, a task fMRI where subjects viewed a fixed set of 480 images. Figure S3 shows the experimental design of the task fMRI. Stimuli were presented for 2s one and 1s off, and were organized into blocks for each condition. 8 unique stimuli were presented per block, with one image repeated in each block for use as a one-back behavioral task. To encourage consistent attention, subjects were instructed to maintain fixation on a central dot, and press a button when they observed the repeated stimulus. A single 350-second scan consisted of 10 27-second stimulus blocks with 6 seconds of rest between blocks. Each session consisted of 7-10 task scans. Stimulus images were square cropped and resized to 8.4° $\times$ 8.4° and presented using a Nordic Neuro Lab 32" LCD monitor positioned at the head of the scanner bed. FMRI data consisted of posterior oblique-axial slices oriented to capture early visual areas and the ventral visual stream (gradient-echo EPI, 2.25x2.25x3.00mm, 27 interleaved slices, TR=1.45s, TE=32ms, phase-encoding in the A>>P direction). EPI susceptibility distortion was estimated using pairs of spin-echo scans with reversed phase-encoding directions \cite{andersson2003correct}. Preprocessing included slice-timing correction with upsampling to 1 second TR, followed by a single-step spatial interpolation combining motion, distortion, and resampling to 2mm isotropic voxels. 

The stimuli used in the task fMRI were 240 natural images selected from the union of all individual-specific images shown to the NSD subjects (9,000 $\times$ 8 = 72,000) and 240 synthetic images created by NeuroGen \cite{gu2022neurogen}, a generative framework that can create synthetic images within a given image category. For the first scan's task fMRI, there were total 4 image conditions for each primary target region (FFA1, EBA, and VWFA1), namely "GroupMaxSyn", "GroupMaxNat", "GroupAvgSyn" and "GroupAvgNat", each containing 40 images (3 regions x 4 conditions x 40 images = 480 images total). The "GroupMaxNat" or "GroupAvgNat" are the natural images that achieve maximal or average predicted activations from the NSD group level encoding model for the region in question, while the "GroupMaxSyn" or "GroupAvgSyn" images are synthetic images optimized using NeuroGen to achieve maximal or average predicted activations from the NSD group level encoding model for the region in question.

During the second MRI scan, the six individuals were shown a set of 128 images (half natural and half synthetic) over 6 conditions designed for FFA1 (32 "SelfMaxSyn", 32 "SelfMaxNat", 20 "OtherMaxSyn", 20 "OtherMaxNat", 12 "GroupMaxSyn" and 12 "GroupMaxNat"). "GroupMaxSyn" and "GroupMaxNat" images are the same as the images in Session 1 for FFA1. "SelfMaxSyn" and "SelfMaxNat" were images from the "Self" personalized linear ensemble encoding model created for that individual (see details in Method \textit{Personalized encoding model construction} section) while "OtherMaxSyn" and "OtherMaxNat" were images from other individuals' "SelfMaxNat" and "SelfMaxSyn" image sets. The fMRI experimental setup and image preprocessing were identical to Session 1. 

\subsection*{Deepnet feature weighted receptive field encoding model}
The encoding model used in this work follows the architecture of the deepnet feature weighted receptive field (deepnet-fwRF) described previously \cite{st2018feature}. The deepnet-fwRF model uses Alexnet \cite{krizhevsky2012imagenet} as a backbone to extract salient features from images. Then the number of features in each AlexNet layer were reduced by selecting those that had the highest variance. A Gaussian pooling field was applied to the feature maps to further reduce the number of features before the final ridge regression which mapped the features to brain regions' responses, which is the average of the voxel-wise activation maps over that region (a scalar). We trained this model for each region and subject in the NSD dataset. The group level model for each region is constructed by averaging the predictions from the eight NSD subjects' regional encoding models.  

\subsection*{Personalized encoding model construction}
We followed our previously developed approach that allows creating personalized linear ensemble models for novel, prospective individuals using small data  \cite{gu2022personalized}. This approach was shown to have a good balance between prediction accuracy and its ability to preserve inter-individual differences in responses. The linear ensemble model linearly combines predictions from a set of base encoding models, which are trained on large data. Here, as in our previous publication, the base models are the deepnet-fwRF \cite{st2018feature} encoding models trained on each NSD subjects' data. To fit the linear ensemble model to predict brain responses in the prospective NeuroGen subjects, we trained on a subset of the Session 1 data consisting of 32 randomly chosen image-response pairs from each image condition and each region (total 32 $\times$ 4 conditions $\times$ 3 regions = 384 images), with the remaining image-response pairs being used to test the personalized encoding model accuracy. 

\subsection*{NeuroGen for optimal image synthesis}
We use here our previously developed NeuroGen framework \cite{gu2022neurogen}, illustrated in Figure S1, which generates images designed to achieve a user-defined brain activation pattern. Essentially, NeuroGen concatenates an image generator (ImageNet pretrained BigGAN-deep \cite{brock2018large}) with an encoding model of human vision. The BigGAN-deep generator takes as input a one-hot encoded class vector and a noise vector, where the class vector indicates the ImageNet 1000 class and the noise vector is initialized with random values and gets updated during optimization. The output of the generator, which is a image, is used as input to the encoding model which then provides the predicted brain responses to that image. By defining the loss function to capture the match between this predicted brain response and the desired brain response, we can iteratively optimize the noise vector to produce the image that minimizes the loss. For the "Max" conditions, the loss function was the negative of the predicted activation plus a regularization term on the noise vector; and for the "Avg" conditions, the loss function was the absolute value of the predicted activation plus a regularization term. The "Group" encoding model (the average of the 8 NSD subjects' deepnet-fwRF encoding models) was used to generate Session 1's images while the personalized encoding models (individual-level linear ensembles) were used to generate Session 2's images. 

\subsection*{Linear mixed effects modeling to test for response differences}
We used a linear mixed effects (LME) model to test for statistical differences in the magnitude of the brain responses to different image conditions. Session 1 comparisons were made for contrasts 1) "GroupAvgSyn" vs "GroupMaxSyn", 2) "GroupAvgNat" vs "GroupMaxNat", 3) "GroupMaxNat" vs "GroupMaxSyn" and 4) "GroupAvgNat" vs "GroupAvgSyn". Session 2 comparisons were made for contrasts 1) "OtherMaxSyn" vs "SelfMaxSyn", 2) "GroupMaxSyn" vs "SelfMaxSyn", 3) "OtherMaxNat" vs "SelfMaxNat", 4) "GroupMaxNat" vs "SelfMaxNat", and 5) "SelfMaxNat" vs "SelfMaxSyn". We assume that there is a population effect of a contrast (across all subjects), but each subject is allowed to have its own random deviation. The LME model is defined as
\begin{equation*}
    \mathbf{y} = \mathbf{X}\mathbf{\beta} + \mathbf{Z}\mathbf{\alpha} + \epsilon
\end{equation*}
where $\mathbf{y}$ is the $6n \times 1$ response vector for the observed fMRI responses to the $n$ images (across both conditions) for all 6 subjects, $\mathbf{X}$ is the binary $6n \times 1$ fixed effects vector containing the image condition information for $n$ images across all 6 subjects, $\beta$ is the fixed-effect coefficient, $\mathbf{Z}$ is the binary $6n \times 6$ random effects matrix containing subject information, $\mathbf{\alpha}$ is the $6 \times 1$ random-effect coefficient vector, and $\epsilon$ is the error in observations. The $p$-values of the model coefficients were calculated using permutation testing, where we randomly permuted the responses from the two image conditions 1000 times and fit the LME model to get $\hat{\beta}$. The $p$-value for the original $\beta$ is the percent of times the random $\hat{\beta}$ is larger in magnitude than the original $\beta$ (two-sided test).

\bibliography{sample}

\begin{thebibliography}{10}
\urlstyle{rm}
\expandafter\ifx\csname url\endcsname\relax
  \def\url#1{\texttt{#1}}\fi
\expandafter\ifx\csname urlprefix\endcsname\relax\def\urlprefix{URL }\fi
\expandafter\ifx\csname doiprefix\endcsname\relax\def\doiprefix{DOI: }\fi
\providecommand{\bibinfo}[2]{#2}
\providecommand{\eprint}[2][]{\url{#2}}

\bibitem{hubel1962receptive}
\bibinfo{author}{Hubel, D.~H.} \& \bibinfo{author}{Wiesel, T.~N.}
\newblock \bibinfo{journal}{\bibinfo{title}{Receptive fields, binocular
  interaction and functional architecture in the cat's visual cortex}}.
\newblock {\emph{\JournalTitle{The Journal of physiology}}}
  \textbf{\bibinfo{volume}{160}}, \bibinfo{pages}{106} (\bibinfo{year}{1962}).

\bibitem{hubel1968receptive}
\bibinfo{author}{Hubel, D.~H.} \& \bibinfo{author}{Wiesel, T.~N.}
\newblock \bibinfo{journal}{\bibinfo{title}{Receptive fields and functional
  architecture of monkey striate cortex}}.
\newblock {\emph{\JournalTitle{The Journal of physiology}}}
  \textbf{\bibinfo{volume}{195}}, \bibinfo{pages}{215--243}
  (\bibinfo{year}{1968}).

\bibitem{kanwisher1997fusiform}
\bibinfo{author}{Kanwisher, N.}, \bibinfo{author}{McDermott, J.} \&
  \bibinfo{author}{Chun, M.~M.}
\newblock \bibinfo{journal}{\bibinfo{title}{The fusiform face area: a module in
  human extrastriate cortex specialized for face perception}}.
\newblock {\emph{\JournalTitle{Journal of neuroscience}}}
  \textbf{\bibinfo{volume}{17}}, \bibinfo{pages}{4302--4311}
  (\bibinfo{year}{1997}).

\bibitem{epstein1998cortical}
\bibinfo{author}{Epstein, R.} \& \bibinfo{author}{Kanwisher, N.}
\newblock \bibinfo{journal}{\bibinfo{title}{A cortical representation of the
  local visual environment}}.
\newblock {\emph{\JournalTitle{Nature}}} \textbf{\bibinfo{volume}{392}},
  \bibinfo{pages}{598--601} (\bibinfo{year}{1998}).

\bibitem{downing2001cortical}
\bibinfo{author}{Downing, P.~E.}, \bibinfo{author}{Jiang, Y.},
  \bibinfo{author}{Shuman, M.} \& \bibinfo{author}{Kanwisher, N.}
\newblock \bibinfo{journal}{\bibinfo{title}{A cortical area selective for
  visual processing of the human body}}.
\newblock {\emph{\JournalTitle{Science}}} \textbf{\bibinfo{volume}{293}},
  \bibinfo{pages}{2470--2473} (\bibinfo{year}{2001}).

\bibitem{kubilius2016deep}
\bibinfo{author}{Kubilius, J.}, \bibinfo{author}{Bracci, S.} \&
  \bibinfo{author}{Op~de Beeck, H.~P.}
\newblock \bibinfo{journal}{\bibinfo{title}{Deep neural networks as a
  computational model for human shape sensitivity}}.
\newblock {\emph{\JournalTitle{PLoS computational biology}}}
  \textbf{\bibinfo{volume}{12}}, \bibinfo{pages}{e1004896}
  (\bibinfo{year}{2016}).

\bibitem{kubilius2019brain}
\bibinfo{author}{Kubilius, J.} \emph{et~al.}
\newblock \bibinfo{journal}{\bibinfo{title}{Brain-like object recognition with
  high-performing shallow recurrent anns}}.
\newblock {\emph{\JournalTitle{Advances in neural information processing
  systems}}} \textbf{\bibinfo{volume}{32}} (\bibinfo{year}{2019}).

\bibitem{zhuang2021unsupervised}
\bibinfo{author}{Zhuang, C.} \emph{et~al.}
\newblock \bibinfo{journal}{\bibinfo{title}{Unsupervised neural network models
  of the ventral visual stream}}.
\newblock {\emph{\JournalTitle{Proceedings of the National Academy of
  Sciences}}} \textbf{\bibinfo{volume}{118}}, \bibinfo{pages}{e2014196118}
  (\bibinfo{year}{2021}).

\bibitem{mehrer2021ecologically}
\bibinfo{author}{Mehrer, J.}, \bibinfo{author}{Spoerer, C.~J.},
  \bibinfo{author}{Jones, E.~C.}, \bibinfo{author}{Kriegeskorte, N.} \&
  \bibinfo{author}{Kietzmann, T.~C.}
\newblock \bibinfo{journal}{\bibinfo{title}{An ecologically motivated image
  dataset for deep learning yields better models of human vision}}.
\newblock {\emph{\JournalTitle{Proceedings of the National Academy of
  Sciences}}} \textbf{\bibinfo{volume}{118}}, \bibinfo{pages}{e2011417118}
  (\bibinfo{year}{2021}).

\bibitem{sexton2022reassessing}
\bibinfo{author}{Sexton, N.~J.} \& \bibinfo{author}{Love, B.~C.}
\newblock \bibinfo{journal}{\bibinfo{title}{Reassessing hierarchical
  correspondences between brain and deep networks through direct interface}}.
\newblock {\emph{\JournalTitle{Science Advances}}}
  \textbf{\bibinfo{volume}{8}}, \bibinfo{pages}{eabm2219}
  (\bibinfo{year}{2022}).

\bibitem{Schrimpf2020integrative}
\bibinfo{author}{Schrimpf, M.} \emph{et~al.}
\newblock \bibinfo{journal}{\bibinfo{title}{Integrative benchmarking to advance
  neurally mechanistic models of human intelligence}}.
\newblock {\emph{\JournalTitle{Neuron}}}  (\bibinfo{year}{2020}).

\bibitem{bowers2022deep}
\bibinfo{author}{Bowers, J.~S.} \emph{et~al.}
\newblock \bibinfo{journal}{\bibinfo{title}{Deep problems with neural network
  models of human vision}}.
\newblock {\emph{\JournalTitle{Behavioral and Brain Sciences}}}
  \bibinfo{pages}{1--74} (\bibinfo{year}{2022}).

\bibitem{brock2018large}
\bibinfo{author}{Brock, A.}, \bibinfo{author}{Donahue, J.} \&
  \bibinfo{author}{Simonyan, K.}
\newblock \bibinfo{title}{Large scale {GAN} training for high fidelity natural
  image synthesis}.
\newblock In \emph{\bibinfo{booktitle}{7th International Conference on Learning
  Representations, {ICLR} 2019, New Orleans, LA, USA, May 6-9, 2019}}
  (\bibinfo{publisher}{OpenReview.net}, \bibinfo{year}{2019}).

\bibitem{van2017neural}
\bibinfo{author}{Van Den~Oord, A.}, \bibinfo{author}{Vinyals, O.} \emph{et~al.}
\newblock \bibinfo{journal}{\bibinfo{title}{Neural discrete representation
  learning}}.
\newblock {\emph{\JournalTitle{Advances in neural information processing
  systems}}} \textbf{\bibinfo{volume}{30}} (\bibinfo{year}{2017}).

\bibitem{rombach2022high}
\bibinfo{author}{Rombach, R.}, \bibinfo{author}{Blattmann, A.},
  \bibinfo{author}{Lorenz, D.}, \bibinfo{author}{Esser, P.} \&
  \bibinfo{author}{Ommer, B.}
\newblock \bibinfo{title}{High-resolution image synthesis with latent diffusion
  models}.
\newblock In \emph{\bibinfo{booktitle}{Proceedings of the IEEE/CVF Conference
  on Computer Vision and Pattern Recognition}}, \bibinfo{pages}{10684--10695}
  (\bibinfo{year}{2022}).

\bibitem{ozcelik2022reconstruction}
\bibinfo{author}{Ozcelik, F.}, \bibinfo{author}{Choksi, B.},
  \bibinfo{author}{Mozafari, M.}, \bibinfo{author}{Reddy, L.} \&
  \bibinfo{author}{VanRullen, R.}
\newblock \bibinfo{journal}{\bibinfo{title}{Reconstruction of perceived images
  from fmri patterns and semantic brain exploration using instance-conditioned
  gans}}.
\newblock {\emph{\JournalTitle{arXiv preprint arXiv:2202.12692}}}
  (\bibinfo{year}{2022}).

\bibitem{gu2022decoding}
\bibinfo{author}{Gu, Z.}, \bibinfo{author}{Jamison, K.},
  \bibinfo{author}{Kuceyeski, A.} \& \bibinfo{author}{Sabuncu, M.}
\newblock \bibinfo{journal}{\bibinfo{title}{Decoding natural image stimuli from
  fmri data with a surface-based convolutional network}}.
\newblock {\emph{\JournalTitle{arXiv preprint arXiv:2212.02409}}}
  (\bibinfo{year}{2022}).

\bibitem{nguyen2016synthesizing}
\bibinfo{author}{Nguyen, A.}, \bibinfo{author}{Dosovitskiy, A.},
  \bibinfo{author}{Yosinski, J.}, \bibinfo{author}{Brox, T.} \&
  \bibinfo{author}{Clune, J.}
\newblock \bibinfo{journal}{\bibinfo{title}{Synthesizing the preferred inputs
  for neurons in neural networks via deep generator networks}}.
\newblock {\emph{\JournalTitle{Advances in neural information processing
  systems}}} \textbf{\bibinfo{volume}{29}} (\bibinfo{year}{2016}).

\bibitem{bashivan2019neural}
\bibinfo{author}{Bashivan, P.}, \bibinfo{author}{Kar, K.} \&
  \bibinfo{author}{DiCarlo, J.~J.}
\newblock \bibinfo{journal}{\bibinfo{title}{Neural population control via deep
  image synthesis}}.
\newblock {\emph{\JournalTitle{Science}}} \textbf{\bibinfo{volume}{364}},
  \bibinfo{pages}{eaav9436} (\bibinfo{year}{2019}).

\bibitem{ponce2019evolving}
\bibinfo{author}{Ponce, C.~R.} \emph{et~al.}
\newblock \bibinfo{journal}{\bibinfo{title}{Evolving images for visual neurons
  using a deep generative network reveals coding principles and neuronal
  preferences}}.
\newblock {\emph{\JournalTitle{Cell}}} \textbf{\bibinfo{volume}{177}},
  \bibinfo{pages}{999--1009} (\bibinfo{year}{2019}).

\bibitem{ratan2021computational}
\bibinfo{author}{Ratan~Murty, N.~A.}, \bibinfo{author}{Bashivan, P.},
  \bibinfo{author}{Abate, A.}, \bibinfo{author}{DiCarlo, J.~J.} \&
  \bibinfo{author}{Kanwisher, N.}
\newblock \bibinfo{journal}{\bibinfo{title}{Computational models of
  category-selective brain regions enable high-throughput tests of
  selectivity}}.
\newblock {\emph{\JournalTitle{Nature communications}}}
  \textbf{\bibinfo{volume}{12}}, \bibinfo{pages}{1--14} (\bibinfo{year}{2021}).

\bibitem{gu2022neurogen}
\bibinfo{author}{Gu, Z.} \emph{et~al.}
\newblock \bibinfo{journal}{\bibinfo{title}{Neurogen: activation optimized
  image synthesis for discovery neuroscience}}.
\newblock {\emph{\JournalTitle{NeuroImage}}} \textbf{\bibinfo{volume}{247}},
  \bibinfo{pages}{118812} (\bibinfo{year}{2022}).

\bibitem{allen2021massive}
\bibinfo{author}{Allen, E.~J.} \emph{et~al.}
\newblock \bibinfo{journal}{\bibinfo{title}{A massive 7t fmri dataset to bridge
  cognitive neuroscience and artificial intelligence}}.
\newblock {\emph{\JournalTitle{Nature neuroscience}}} \bibinfo{pages}{1--11}
  (\bibinfo{year}{2021}).

\bibitem{gu2022personalized}
\bibinfo{author}{Gu, Z.}, \bibinfo{author}{Jamison, K.},
  \bibinfo{author}{Sabuncu, M.} \& \bibinfo{author}{Kuceyeski, A.}
\newblock \bibinfo{journal}{\bibinfo{title}{Personalized visual encoding model
  construction with small data}}.
\newblock {\emph{\JournalTitle{arXiv preprint arXiv:2202.02245}}}
  (\bibinfo{year}{2022}).

\bibitem{st2018feature}
\bibinfo{author}{St-Yves, G.} \& \bibinfo{author}{Naselaris, T.}
\newblock \bibinfo{journal}{\bibinfo{title}{The feature-weighted receptive
  field: an interpretable encoding model for complex feature spaces}}.
\newblock {\emph{\JournalTitle{NeuroImage}}} \textbf{\bibinfo{volume}{180}},
  \bibinfo{pages}{188--202} (\bibinfo{year}{2018}).

\bibitem{saxe2006divide}
\bibinfo{author}{Saxe, R.}, \bibinfo{author}{Brett, M.} \&
  \bibinfo{author}{Kanwisher, N.}
\newblock \bibinfo{journal}{\bibinfo{title}{Divide and conquer: a defense of
  functional localizers}}.
\newblock {\emph{\JournalTitle{Neuroimage}}} \textbf{\bibinfo{volume}{30}},
  \bibinfo{pages}{1088--1096} (\bibinfo{year}{2006}).

\bibitem{friston2006critique}
\bibinfo{author}{Friston, K.~J.}, \bibinfo{author}{Rotshtein, P.},
  \bibinfo{author}{Geng, J.~J.}, \bibinfo{author}{Sterzer, P.} \&
  \bibinfo{author}{Henson, R.~N.}
\newblock \bibinfo{journal}{\bibinfo{title}{A critique of functional
  localisers}}.
\newblock {\emph{\JournalTitle{Neuroimage}}} \textbf{\bibinfo{volume}{30}},
  \bibinfo{pages}{1077--1087} (\bibinfo{year}{2006}).

\bibitem{grill2001lateral}
\bibinfo{author}{Grill-Spector, K.}, \bibinfo{author}{Kourtzi, Z.} \&
  \bibinfo{author}{Kanwisher, N.}
\newblock \bibinfo{journal}{\bibinfo{title}{The lateral occipital complex and
  its role in object recognition}}.
\newblock {\emph{\JournalTitle{Vision research}}}
  \textbf{\bibinfo{volume}{41}}, \bibinfo{pages}{1409--1422}
  (\bibinfo{year}{2001}).

\bibitem{cohen2002language}
\bibinfo{author}{Cohen, L.} \emph{et~al.}
\newblock \bibinfo{journal}{\bibinfo{title}{Language-specific tuning of visual
  cortex? functional properties of the visual word form area}}.
\newblock {\emph{\JournalTitle{Brain}}} \textbf{\bibinfo{volume}{125}},
  \bibinfo{pages}{1054--1069} (\bibinfo{year}{2002}).

\bibitem{collins2014beyond}
\bibinfo{author}{Collins, J.~A.} \& \bibinfo{author}{Olson, I.~R.}
\newblock \bibinfo{journal}{\bibinfo{title}{Beyond the ffa: the role of the
  ventral anterior temporal lobes in face processing}}.
\newblock {\emph{\JournalTitle{Neuropsychologia}}}
  \textbf{\bibinfo{volume}{61}}, \bibinfo{pages}{65--79}
  (\bibinfo{year}{2014}).

\bibitem{henriksson2015faciotopy}
\bibinfo{author}{Henriksson, L.}, \bibinfo{author}{Mur, M.} \&
  \bibinfo{author}{Kriegeskorte, N.}
\newblock \bibinfo{journal}{\bibinfo{title}{Faciotopy—a face-feature map with
  face-like topology in the human occipital face area}}.
\newblock {\emph{\JournalTitle{Cortex}}} \textbf{\bibinfo{volume}{72}},
  \bibinfo{pages}{156--167} (\bibinfo{year}{2015}).

\bibitem{barense2010medial}
\bibinfo{author}{Barense, M.~D.}, \bibinfo{author}{Henson, R.~N.},
  \bibinfo{author}{Lee, A.~C.} \& \bibinfo{author}{Graham, K.~S.}
\newblock \bibinfo{journal}{\bibinfo{title}{Medial temporal lobe activity
  during complex discrimination of faces, objects, and scenes: Effects of
  viewpoint}}.
\newblock {\emph{\JournalTitle{Hippocampus}}} \textbf{\bibinfo{volume}{20}},
  \bibinfo{pages}{389--401} (\bibinfo{year}{2010}).

\bibitem{yang2016anterior}
\bibinfo{author}{Yang, H.}, \bibinfo{author}{Susilo, T.} \&
  \bibinfo{author}{Duchaine, B.}
\newblock \bibinfo{journal}{\bibinfo{title}{The anterior temporal face area
  contains invariant representations of face identity that can persist despite
  the loss of right ffa and ofa}}.
\newblock {\emph{\JournalTitle{Cerebral Cortex}}}
  \textbf{\bibinfo{volume}{26}}, \bibinfo{pages}{1096--1107}
  (\bibinfo{year}{2016}).

\bibitem{midjourney}
\bibinfo{title}{Midjourney}.
\newblock \bibinfo{howpublished}{\url{http://www.midjourney.com}}.

\bibitem{lin2014microsoft}
\bibinfo{author}{Lin, T.-Y.} \emph{et~al.}
\newblock \bibinfo{title}{Microsoft coco: Common objects in context}.
\newblock In \emph{\bibinfo{booktitle}{European conference on computer
  vision}}, \bibinfo{pages}{740--755} (\bibinfo{organization}{Springer},
  \bibinfo{year}{2014}).

\bibitem{charest2018glmdenoise}
\bibinfo{author}{Charest, I.}, \bibinfo{author}{Kriegeskorte, N.} \&
  \bibinfo{author}{Kay, K.~N.}
\newblock \bibinfo{journal}{\bibinfo{title}{Glmdenoise improves multivariate
  pattern analysis of fmri data}}.
\newblock {\emph{\JournalTitle{NeuroImage}}} \textbf{\bibinfo{volume}{183}},
  \bibinfo{pages}{606--616} (\bibinfo{year}{2018}).

\bibitem{kay2013glmdenoise}
\bibinfo{author}{Kay, K.}, \bibinfo{author}{Rokem, A.},
  \bibinfo{author}{Winawer, J.}, \bibinfo{author}{Dougherty, R.} \&
  \bibinfo{author}{Wandell, B.}
\newblock \bibinfo{journal}{\bibinfo{title}{Glmdenoise: a fast, automated
  technique for denoising task-based fmri data}}.
\newblock {\emph{\JournalTitle{Frontiers in neuroscience}}}
  \textbf{\bibinfo{volume}{7}}, \bibinfo{pages}{247} (\bibinfo{year}{2013}).

\bibitem{prince2022improving}
\bibinfo{author}{Prince, J.~S.} \emph{et~al.}
\newblock \bibinfo{journal}{\bibinfo{title}{Improving the accuracy of
  single-trial fmri response estimates using glmsingle}}.
\newblock {\emph{\JournalTitle{Elife}}} \textbf{\bibinfo{volume}{11}},
  \bibinfo{pages}{e77599} (\bibinfo{year}{2022}).

\bibitem{rokem2020fractional}
\bibinfo{author}{Rokem, A.} \& \bibinfo{author}{Kay, K.}
\newblock \bibinfo{journal}{\bibinfo{title}{Fractional ridge regression: a
  fast, interpretable reparameterization of ridge regression}}.
\newblock {\emph{\JournalTitle{GigaScience}}} \textbf{\bibinfo{volume}{9}},
  \bibinfo{pages}{giaa133} (\bibinfo{year}{2020}).

\bibitem{stigliani2015temporal}
\bibinfo{author}{Stigliani, A.}, \bibinfo{author}{Weiner, K.~S.} \&
  \bibinfo{author}{Grill-Spector, K.}
\newblock \bibinfo{journal}{\bibinfo{title}{Temporal processing capacity in
  high-level visual cortex is domain specific}}.
\newblock {\emph{\JournalTitle{Journal of Neuroscience}}}
  \textbf{\bibinfo{volume}{35}}, \bibinfo{pages}{12412--12424}
  (\bibinfo{year}{2015}).

\bibitem{andersson2003correct}
\bibinfo{author}{Andersson, J.~L.}, \bibinfo{author}{Skare, S.} \&
  \bibinfo{author}{Ashburner, J.}
\newblock \bibinfo{journal}{\bibinfo{title}{How to correct susceptibility
  distortions in spin-echo echo-planar images: application to diffusion tensor
  imaging}}.
\newblock {\emph{\JournalTitle{Neuroimage}}} \textbf{\bibinfo{volume}{20}},
  \bibinfo{pages}{870--888} (\bibinfo{year}{2003}).

\bibitem{krizhevsky2012imagenet}
\bibinfo{author}{Krizhevsky, A.}, \bibinfo{author}{Sutskever, I.} \&
  \bibinfo{author}{Hinton, G.~E.}
\newblock \bibinfo{journal}{\bibinfo{title}{Imagenet classification with deep
  convolutional neural networks}}.
\newblock {\emph{\JournalTitle{Advances in neural information processing
  systems}}} \textbf{\bibinfo{volume}{25}}, \bibinfo{pages}{1097--1105}
  (\bibinfo{year}{2012}).

\bibitem{dworkin2020extent}
\bibinfo{author}{Dworkin, J.~D.} \emph{et~al.}
\newblock \bibinfo{journal}{\bibinfo{title}{The extent and drivers of gender
  imbalance in neuroscience reference lists}}.
\newblock {\emph{\JournalTitle{Nature Neuroscience}}}
  \textbf{\bibinfo{volume}{23}}, \bibinfo{pages}{918--926}
  (\bibinfo{year}{2020}).

\end{thebibliography}

\section*{Acknowledgements}
This work was funded by the following grants: R01 NS102646 (AK), RF1 MH123232 (AK), R01 LM012719 (MS), R01 AG053949 (MS), NSF CAREER 1748377 (MS), NSF NeuroNex Grant 1707312 (MS), and Cornell/Weill Cornell Intercampus Pilot Grant (AK and MS). The NSD data were collected by Kendrick Kay and Thomas Naselaris under the NSF CRCNS grants IIS-1822683 and IIS-1822929.

\section*{Author contributions statement}
A.K., M.S., Z.G. and K.J. conceived and conducted the experiments and interpreted the results,  Z.G. additionally analysed the results and carried out the statistical analyses. K.J. additionally processed the imaging data and designed the MRI experiment protocols. Z.G. and A.K. wrote the manuscript. All authors reviewed the manuscript.

\section*{Citation gender diversity statement}
Recent work in several fields of science has identified a bias in citation practices such that papers from women and other minorities are under-cited relative to the number of such papers in the field \cite{dworkin2020extent}. Here we sought to proactively consider choosing references that reflect the diversity of the field in thought, form of contribution, gender, and other factors. We obtained predicted gender of the first and last author of each reference by using databases that store the probability of a name being carried by a woman \cite{dworkin2020extent}. By this measure (and excluding self-citations to the first and last authors of our current paper),  our references contain 8.61\% woman(first)/woman(last), 18.71\% man/woman, 7.89\% woman/man, and 64.79\% man/man. This method is limited in that a) names, pronouns, and social media profiles used to construct the databases may not, in every case, be indicative of gender identity and b) it cannot account for intersex, non-binary, or transgender people. We look forward to future work that could help us to better understand how to support equitable practices in science.

\end{document}